# Generation of Intense High-Order Vortex Harmonics

Xiaomei Zhang, Baifei Shen[*], Yin Shi, Xiaofeng Wang, Lingang Zhang, Wenpeng Wang, Jiancai Xu, Longqiong Yi, and Zhizhan Xu

*State Key Laboratory of High Field Laser Physics, Shanghai Institute of Optics and Fine Mechanics, Chinese Academy of Sciences, Shanghai 201800, China*

**Abstract:**

This paper presents the method for the first time to generate intense high-order optical vortices that carry orbital angular momentum in the extreme ultraviolet region. In three-dimensional particle-in-cell simulation, both the reflected and transmitted light beams include high-order harmonics of the Laguerre-Gaussian (LG) mode when a linearly polarized LG laser pulse impinges on a solid foil. The mode of the generated LG harmonic scales with its order, in good agreement with our theoretical analysis. The intensity of the generated high-order vortex harmonics is close to the relativistic region, and the pulse duration can be in attosecond scale. The obtained intense vortex beam possesses the combined properties of fine transversal structure due to the high-order mode and the fine longitudinal structure due to the short wavelength of the high-order harmonics. Thus, the obtained intense vortex beam may have extraordinarily promising applications for high-capacity quantum information and for high-resolution detection in both spatial and temporal scales because of the addition of a new degree of freedom.

[*] Author to whom correspondence should be addressed. Electronic mail: bfshen@mail.shcnc.ac.cn.

**Text:**

Light beams can exhibit helical wave fronts: the light phase "winds up" around the spatial beam center and forms an optical vortex. The phase wind imprints an orbital angular momentum (OAM) to the beam [1, 2]. The characteristic helical phase profiles of optical vortices are described by $\exp(il\phi)$ multipliers, where $\phi$ is the azimuthal coordinate and the integer number $l$ is their topological charge, corresponding the order of the mode. The total phase accumulated in one full annular loop is $2\pi l$, and an OAM of $l\hbar$ is carried by per photon for an $l$-order linearly polarized optical vortex beam. Based on this, the high order optical vortex beam provides as a powerful tool in quantum information to investigate the entanglement state [3] and for studies of cold atoms and enhancing atomic transition [4-7].

In order to provide more quantum information and for other potential applications, high order vortex beams are required. However, limited by the etching resolution, the common method using forked diffraction grating [8] or the spiral phase plates [1] to generate the optical vortex beams is difficult to be used to obtain them. Many studies have attempted to generate light beams with OAM. For example, a relativistic electron beam can act as a mode converter that interacts with a laser in a helical undulator [9-11] and high-energy photons in MeV–GeV with OAM can be obtained by Compton Backscattering of twisted laser photons off relativistic electrons [12, 13], where the mode of the Laguerre–Gaussian (LG) pulse remains unchanged. In addition, in view of the gas high-order harmonics generation (HHG) scheme [14-16], the observed harmonics possess a helical wave-front in both experimental [17, 18] and theoretical studies [19, 20] when a $\sim 10^{15} W/cm^2$ helical beam is focused into a gas jet.

While those interesting findings are expected to be studied more in the future, in this Letter, we present the first way to generate intense high-order helical beams in the extreme ultraviolet (XUV) region using plasma as the mode-converter. When an intense linearly polarized driving vortex beam with low-order mode impinges on a solid target, both the reflected and transmitted light include high-order harmonics, the phase of which scales with the harmonic order. In this nonlinear process, OAM carried by the driving helical light is readily transferred to the harmonics. The mode of the generated optical vertex beam, can easily reach over ten or even a few tens and keeps well when it propagates. Moreover, plasma as the nonlinear medium shows superiority in the damage threshold compared with the optical component. With the present method, the intensity of the generated high-order vortices is close to the relativistic region. In addition, such generated beam is ultra-short (attosecond) in time scale, making it an ideal tool for probing electronic dynamics on the atomic or molecular scale. Our method may for the first time produce LG pulses simultaneously of high intensity, short wavelength, and high order mode.

The proposed scheme is confirmed with three-dimensional (3D) particle-in-cell (PIC) simulations based on the VORPAL code. The driving LG beam is *p*-polarized, described as

$$a(LG_{lp}) = a_0 \left(\frac{\sqrt{2}r}{r_0}\right)^l \exp\left(-\frac{r^2}{r_0^2}\right) \exp(il\phi)(-1)^p L_p^l\left(\frac{2r^2}{r_0^2}\right) \sin^2\left(\frac{\pi t}{2t_0}\right). \quad (1)$$

In this simulation, $l=1$, $p=0$ meaning $LG_{10}$ beam (low order optical vortex mode) is used, $r_0 = 6$ μm and $t_0 = 3T$, where $T$ is the driving laser period and $\phi$ is the azimuthal coordinate (relating to the positon of *y*, *z*) within the range of $[0\ 2\pi]$. For the laser wavelength $\lambda_0 = 0.8$ μm, $a_0 = eA/m_e c^2 = 2$, corresponds a peak electric field intensity $4\times10^{12}$ V/cm, where *A* is the vector potential, *c* is the light speed in vacuum, $m_e$ is the electron mass, and *e* is the electron charge. The thin foil as the nonlinear converter occupies the region $15$ μm $< x < 16$ μm in the propagation direction of the driving beam and $-12$ μm $< y(z) < 12$ μm in the transverse direction with a density of $n_0 = 1\times10^{22}$ /cm$^3$. The simulation box is $20$ μm(*x*)$\times 24$ μm(*y*)$\times 24$ μm(*z*), which corresponds to a window with $1000\times400\times400$ cells and two particles per cell. At $t=0$, the laser pulse enters the simulation box from the left boundary.

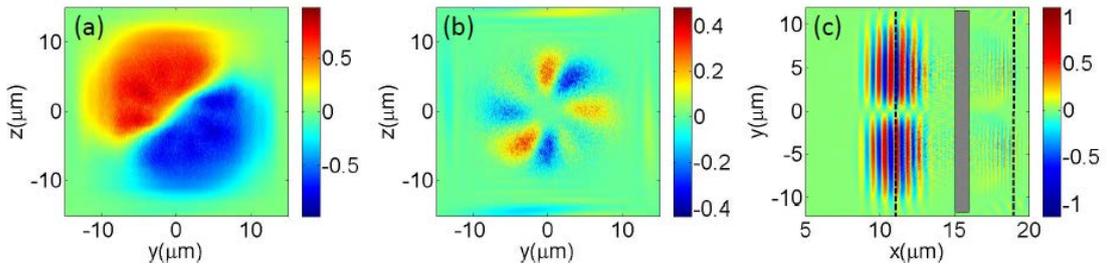

Fig. 1. (a) Electric field $E_y$ of the reflected pulse in the *z-y* plane at $x = 11$ μm, (b) electric field $E_y$ of the transmitted pulse in the *z-y* plane at $x = 19$ μm and (c) total electric field $E_y$ in the *x-y* plane

at $z = 0$ μm at $t = 74.7$ fs when the incident $LG_{10}$ beam is completely reflected from the solid target. The black dotted lines in (c) show the $x$ positions of planes in (a) and (b), and the grey box shows the foil position. The field is normalized to $m_e\omega_0 c/e$ ($4\times10^{12}$ V/m).

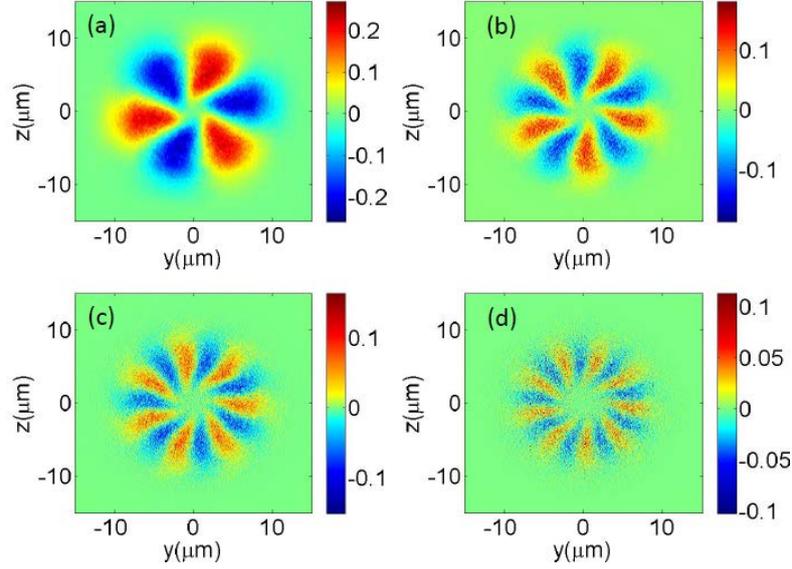

Fig. 2. (a) Reflected electric field distribution of the third, fifth, seventh, and ninth harmonics in the $z$-$y$ plane at $x = 11$ μm at the same time as that in Fig. 1. More information about the third harmonic propagating is available in the Supplemental Material (the plane in Fig. 1(a). changing with the $x$ position from $x = 10.8$ μm to $x = 11.54$ μm with $\Delta x = 0.02$ μm).

Fig.1 shows the electric field distributions of the reflected and transmitted pulses in the $z$-$y$ plane when the driving beam is completely reflected from the solid foil [see Fig.1(c)]. A comparison of Figs. 1(a) and (b) shows that the mode of the transmitted field has a complicated structure that may contain high-order LG modes, whereas the transverse plane of the reflected field still shows the $LG_{10}$ mode like the driving field. To see the information of harmonics modes more clearly, we plot the transverse electric field distributions of the third, fifth, seventh, and ninth harmonics of the reflected field in Fig. 2, and the harmonic mode can then be seen clearly. According to the number of inter-wined helices, the corresponding modes of these harmonics are $l$=3, 5, 7, 9 respectively. Supplemental Material (reflected electric field distribution of the third harmonic in the $z$-$y$ plane by changing $x$ positon from $x = 10.8$ μm to $x = 11.54$ μm with $\Delta x = 0.02$ μm) clearly and intuitively shows the helical feature. The distance of rotating one loop of $E_y$ is approximately $\lambda_3 = 0.2667$ μm, which is just equal to the wavelength of the third harmonics $\lambda_0/3$. Seeing from the frequency spectrum, we know that the expected harmonics information in Fig. 1(a) is covered because of the much higher intensity of the fundamental field, that is, the $LG_{10}$ mode. For the transmitted beam, the fundamental field is completely reflected by the high-density foil. Thus, the complicated structure in the *y-z* plane can be shown.

When an intense linearly polarized laser pulse imprints on a solid foil, the foil surface oscillates with double frequency of the incident pulse since the pondermotive force it feels is $\propto (1-\cos(2\omega_0 t))$, where $\omega_0$ is the incident laser frequency. HHG in both reflected and transmitted beams occurs when the driving beam is a fundamental Gaussian pulse, that is, $LG_{00}$ mode, which has been elucidated clearly with the simple well-known oscillating mirror model [21-27]. However, in the present case, the oscillating mirror driven by the low-order vortex beam becomes a "vortex oscillating mirror" (VOM). On one side, the mirror is oscillating in the longitudinal direction; on the other side, it is helical in azimuthal direction transferred from the driving beam. The phase of harmonics radiated from this VOM changes accordingly. The reflected field can be expressed by

$$E \sim a_0 \sin\left(\omega_0 t + l\phi + \kappa \sin\left(2(\omega_0 t + l\phi)\right)\right), \quad (2)$$

where $\kappa$ is related to the oscillating amplitude. After the Fourier expansion of Eq. (1), we can obtain

$$E/a_0 \sim \sum_{n=0}^{\infty} J_n(\kappa) \sin\left((2n+1)(\omega_0 t + l\phi)\right), \quad (3)$$

where $J_n$ is the first kind Bessel function. From Eq.(2), we can see that reflected field indeed includes odd order harmonics just like previous studies of solid HHG. However, the most important is the phase of the harmonic scales with its harmonic order. The topological charge of the *m*-order harmonic is $lm$. For example, the LG mode of the third harmonic should be $LG_{30}$ when $l=1$. In another point of view, in the nonlinear process of HHG, a photon of the *m*-order harmonic is transformed from *m* photons of the fundamental light carrying the OAM of $\hbar$ by per photon. Therefore, the OAM of $m\hbar$ per photon is carried for the *m*-order harmonic in terms of energy and momentum conservations.

According to the definition of the $lm$ order LG mode, the phase changes $lm$ times from 0 to $2\pi$ in one annular loop in a fixed transverse plane, that is, one annular loop contains $2lm$ intensity (field) peaks. Therefore, the angle between the two adjacent intensity peaks is $\pi/lm$, and the angle differences between the intensity peaks of the third and fifth orders, fifth and seventh orders, seventh and ninth orders, and $lm$ order and $(lm+2)$ order in the transverse intensity plane are $2\pi/15$, $2\pi/35$, $2\pi/63$ and $2\pi/lm(lm+2)$, respectively. This result agrees with the display in Fig. 2.

Fig. 3(a) shows the intensity profiles of the different harmonics in the transverse direction, which are also the intensity profiles of their corresponding modes. The typical donut distribution of the $LG_{l0}$ mode can be clearly observed. For example, the $LG_{30}$ mode of the third harmonic beam is shown in Fig. 2(b). Here, we should note the obtained intensity radius in the transverse direction for different modes keeps same with that of the driving pulse, shown in Fig. 3(a), which may be slightly different from the standard high order LG mode. Fig. 3(c) gives the scaling of the harmonic spectrum with intensity and the conversion efficiency of different modes (harmonics). The spectrum rolloff is fitted by $I_\omega \sim \omega^{-2.98}$, which agrees well with that from the oscillating mirror model or the improved "spiky" mirror model [25]. In the present study, the energy conversion efficiency from the driving $LG_{10}$ pulse to the high-order mode LG pulse is as high as that of the general solid HHG mechanism. For example, approximately 3% energy of driving pulse is transferred to the $LG_{30}$ mode. This energy is expected to be increased with increasing driving pulse intensity.

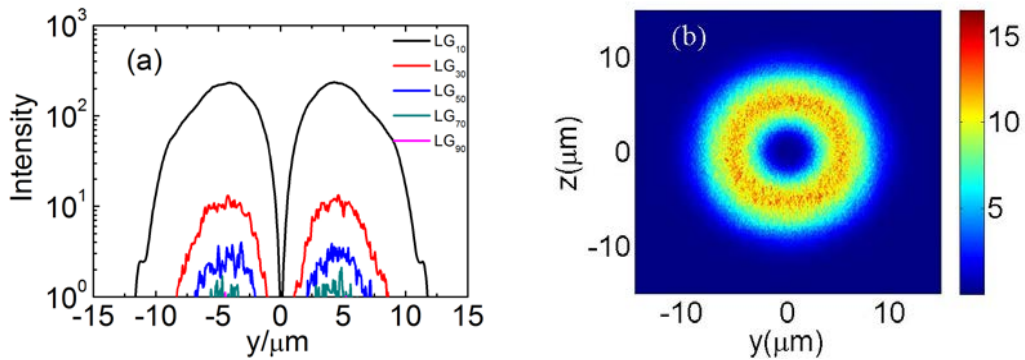

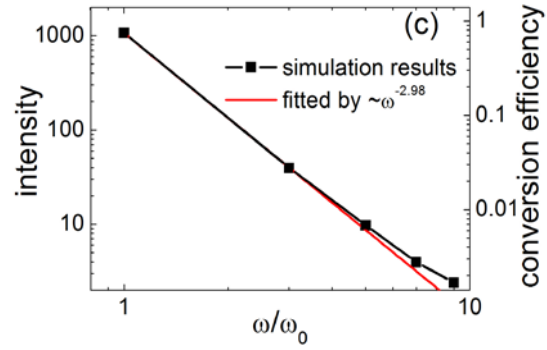

Fig. 3. (a) Transverse intensity profiles of the fundamental, third, fifth, seventh, and ninth harmonics with the position $y(z)$. (b) Transverse intensity distribution of the third harmonics of $LG_{30}$ mode. (c) Harmonic spectrum and conversion efficiency distributions of different modes. The intensity in (a-c) is normalized to $(m_e\omega_0 c/e)^2$ and the time is the same as that in Fig. 1.

In spatial scale, we can obtain intense high-order optical vertices, such as $LG_{30}$, $LG_{50}$, $LG_{70}$, and $LG_{90}$ beam, and even higher order vertices with fine transversal structures. Conversely, the obtained intense vortex beam may have a fine longitudinal structure in temporal scale when plasma is used as the ideal nonlinear medium to generate high-order harmonics. As shown in Fig. 4, the helical attosecond pulse (~100 as) train is obtained. These helical attosecond structures may be powerful tools in high-resolution indispensable fields. With the combined characteristics of ultrashort pulse and OAM, these intense attosecond XUV optical vortices will also open a new view for applications in particle micro-manipulation imaging [28], optical communication [29, 30], quantum information [31], and so on.

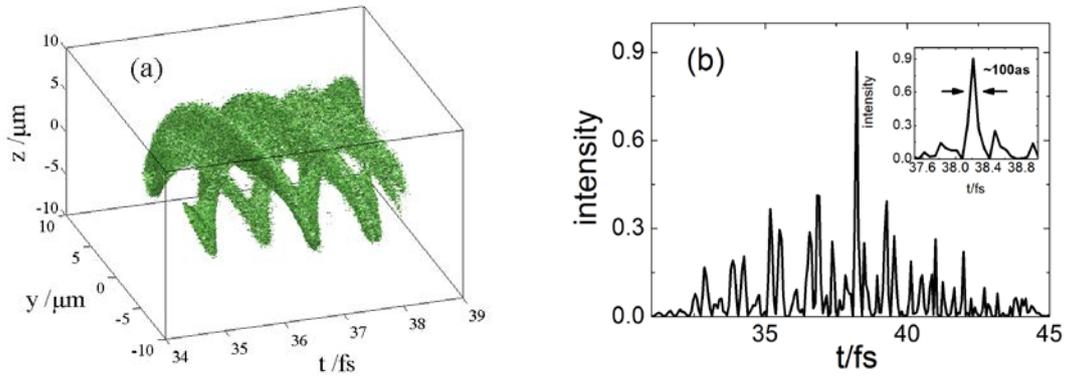

Fig. 4. (a) Isosurface distribution of the high harmonic field after the fundamental part has been filtered with isosurface value $0.025 m_e \omega_0 c / e$ and (b) its detailed attosecond train structure at the azimuthal angle $\phi = 0°$ ($y = 5$ μm, $z = 0$ μm). The intensity in (b) is normalized to $(m_e \omega_0 c / e)^2$ and the time is the same as that in Fig. 1.

The scaling relation between the mode and the order of the harmonics agrees well with the above analysis result. According to the relation in Eq. (2), it is possible to get high order vortices carrying large OAM in the XUV region. The OAM carried by per photon is $lm\hbar$, which can make a huge mechanical torque when it interacts with matter. The huge torque is expected to be applied widely, just like the ponderomotive force of relativistic laser pulse [32]. In the present mode, the red-shifting induced by the surface moving ahead into the propagating direction is not considered. We know the driving laser intensity can be scaled to $10^{21}$ W/cm$^2$ [33] using plasma as the nonlinear medium to generate ultrashort pulses. If the laser pulse is ultra-relativistic, the red-shifting effect is heavy and should be included. That complicated case is beyond of our consideration in this Letter.

In conclusion, we presented a novel method to obtain intense high-order vortices in the high frequency region by irradiating a vortex pulse with a low mode on a solid foil. Different from the generation methods with conventional mode transmission related with optical glass which has a low damage threshold and the nonlinear process in gas driven requiring an appropriate (~ $10^{14-15}$ W/cm$^2$) intensity laser pulse, the proposed method uses the plasma as the nonlinear medium, which can bear relativistic laser intensity. VOM is formed when the solid plasma is irradiated by an optical vortex beam and radiates high harmonics carrying large OAM with high efficiency. The generated high-order optical vortices exert a strong coherent effect, and the mode of the harmonic scales with the harmonic order and the modes of harmonics (LG$_{30}$, LG$_{50}$, LG$_{70}$, and LG$_{90}$) keep well when it propagates. With this method, other order (odd and even order) vertices may be obtained if the driving beam irradiates obliquely or with other arbitrary topological charge. Given the combined the properties of OAM and HHG, the obtained intense vortex beam shows extraordinarily applications in both temporal and spatial scales. The intense light beam carrying large OAM is pivotal for its applications in wide and deep fields since a new degree of freedom is added. Moreover, no obstacle is encountered when up-shifting the generated twisted beams to extremely short and more intense domain because the phase twist is transferred from the driving field with low-order mode. And fortunately, a potential approach to generate intense low-order $LG_{10}$ pulses has been proposed by reflection from structured plasma of a relativistic fundamental laser pulse, which is called the light fan regime [32] and such fundamental twisted light can be successfully applied successfully in particle acceleration [34, 35].

This work was supported by the Ministry of Science and Technology (Grant Nos. 2011DFA11300, and 2011CB808104,) and the National Natural Science Foundation of China (Grant Nos. 61221064, 11374319, 11125526, 11335013, and 11127901).